\documentstyle[11pt,conf_iap,epsf]{article}
\def\plottwo#1#2{\centering \leavevmode
a)\epsfxsize=.45\columnwidth \epsfbox{#1} \hfil
b)\epsfxsize=.45\columnwidth \epsfbox{#2}}
\begin{document}
\begin{flushright}
12$th$ IAP Astrophysics Colloquium, July 1996, Paris, ed. R.Ferlet
\end{flushright}
\heading{THE CHROMATICITY OF MICROLENSING}

\bigskip\bigskip\bigskip
\bigskip\bigskip\bigskip
\bigskip\bigskip\bigskip
\bigskip\bigskip\bigskip

\author{D. D. SASSELOV}
       {Harvard-Smithsonian Center for Astrophysics, Cambridge, 
       MA 02138, USA; dsasselov@cfa.harvard.edu}

\bigskip

\begin{abstract}{\baselineskip 0.4cm 
Gravitational microlensing is not achromatic when the star being microlensed
is resolved by the lens. We show how narrow-band photometry and moderate- to
high-resolution spectroscopy can be used to reconstruct the stellar surface
intensity distribution of a microlensed star. 
Such microlens imaging can provide
a unique opportunity to study the surfaces of normal red giants where
Doppler imaging is not applicable due to slow rotation.}
\end{abstract}

\section{Introduction}
Microlensing events have often been advertised as achromatic, but there are
a number of conditions which will introduce a wavelength dependence in
the light curves and in spectral features of an observed event. Among them
are light from the lens which has a different color from the source
\cite{kam} and blended sources, where only one star of a blend is
likely to be lensed \cite{ala,esi}.
Here we are interested in a third source of chromaticity $-$ 
finite-size
effects, because these arise from resolved features on the surface of the 
lensed star. Such effects are observable at a level at which they provide
an excellent opportunity for stellar surface imaging \cite{sas}.

Finite-size effects have been studied as methods to partially
remove the degeneracy of microlensing light curves through the alterations
of the standard light curve \cite{ag4,ag5,nem,wm,gw},
its polarized emission \cite{sim},
spectral shifts due to stellar rotation \cite{mg},
and narrow-band photometry in resonance lines \cite{ls}.
Here we want to put the emphasis on the inverse problem $-$ reconstructing
the stellar surface features and probing the stellar atmosphere.

The total flux observed is wavelength dependent through its dependence on
$B(r,\theta)$ $-$ the resolved stellar surface intensity distribution, which can
vary strongly with wavelength in selected spectral regions (in continua
and within spectral lines). The reconstruction of $B(r,\theta)$ distinguishes
our approach from the deconvolution of quasar structure from microlensing
light curves (see \cite{gri}).

\section{Probing Stellar Atmospheres}
Stellar research is about to experience a dramatic growth, powered by new
fundamental data of unprecedented quality. One source is the $HIPPARCOS$
satellite with positions and parallaxes; the other source is the
development of stellar interferometers and arrays able to resolve surfaces
of nearby stars in ways achievable only for the Sun until now. 

The Sun offers an instructive example. On one hand, we have measured the
solar radius to 0.001\%, its luminosity to 0.003\% and its surface temperature
to about 2K. On the other hand, the solar disk brightness distribution as a
function of wavelength has been mapped into the depth distribution of
temperature and density, which allows a complete synthesis of the observed
solar spectrum \cite{ku0}. The situation with stars has been confined to very much
lower accuracy. The angular diameters of a very small number of stars have been
measured presently; the accuracy is about 10\% at best \cite{arm}.
Stellar disk brightness distributions (limb darkening)
have been measured only in a handful of stars at the level of a proof-of-concept
(directly $-$ \cite{moz,ste}; or in binaries $-$ \cite{and};
they cannot be used to build a model atmosphere, as in the
case of the Sun. The possible presence of
surface features (like solar spots, active
regions, etc.) is only inferred spectroscopically. Therefore our understanding
of stellar light is implicitly tied to the solar atmosphere model.

This situation is unfortunate for a number of reasons. There has been an
increased demand for accurate stellar models from a number of fields. To
name a few, much improved color-temperature calibrations are needed for
calibrating distance indicators and determination of ages; stellar population
syntheses (e.g. for high-$z$ systems) rely critically on the accuracy of
current stellar models. Our Sun is not a good zero point for most of these
applications.

The advent of large interferometric arrays will go a long way towards solving
these problems, but all of these complex and expensive facilities 
are still on the drawing boards. In the meantime, gravitational microlensing
offers an easily accessible, immediate, and inexpensive way to image at
least some types of stars. It also offers, by its nature, access to stellar
populations (in the Galactic bulge and Magellanic Clouds), which are beyond
the reach of any interferometer.

\section{Practical considerations \& a model for MACHO Alert 95-30}
The amplification of a point source by 
a point mass, $M$, depends only on their
projected separation $d$~, $A(d)= {(d^2+2)/(d(d^2+4)^{1/2})}$,
where $d$ is expressed in units of the angular radius of the
Einstein ring of the 
lens, $\theta_{_E}=([4GM/c^2][D_{_{\rm LS}}/D_{_{\rm OL}}
D_{_{\rm OS}}])^{1/2}$, and $D_{_{\rm OL}}, D_{_{\rm LS}}$,
and $D_{_{\rm OS}}$ are the distances between the observer, lens 
and source. The total flux received from an extended
source is therefore obtained by integration over its 
infinitesimal elements,
$$
F(t)=\int_0^{\tilde{R_s}} r dr B(r) \int_0^{2\pi} d\theta A(d),
$$
where $B(r)$ is the surface brightness profile
of the source in the projected polar coordinates $(r,\theta)$
around its center. 
All projected length scales
are normalized by the Einstein ring radius.
As a result, although gravitational microlensing
is achromatic, when the source is resolved, $F(t)$ is wavelength
dependent through $B(r)$.

Microlensing events towards the bulge occur when a 
low-mass star from the disk (the~lens at D$_{\rm OL}\approx 6 kpc$) 
crosses
the line-of-sight towards a star from the bulge (the source at
D$_{\rm OS}\approx 8 kpc$); typical durations are between weeks and months.
While most lenses are point masses, many ($\sim$ 20\%) of the sources
monitored in the Galactic bulge are red giants and supergiants $-$
their angular radii are {\em comparable} to the Einstein
radius of a sub-solar mass lens (in the range of 50$\mu$as).
In addition to being resolved by most lenses, the projected disks of
bulge giants are large enough to make the probability for a lens transit
very high $-$ in fact, at least one such event (MACHO Alert 95-30) was
very well observed last summer \cite{pra}.   

The conditions for microlens imaging towards the LMC are not as favorable.
However, a large number of lenses are binaries (increasingly more are being
recognized and observed presently), thus caustic crossings should ensure
the necessary resolution. This applies equally to bulge events.
In summary, microlens imaging is viable; below we discuss different
stellar surface features and the observational requirements for their
detection.

As a way to illustrate the opportunities for stellar imaging and the
microlensing computer code described here by \cite{hey}, we
will use our model for the red giant which was the source in MACHO Alert 95-30.
The giant is best fit with a model atmosphere of $T_{\rm eff}=3500K \pm 200K$,
$log~g=1.0 \pm 0.8$, and $V_{\rm t}=2~km s^{-1}$.
The atmosphere is in hydrostatic and radiative equilibrium and has solar
(Pop.I) abundances. The opacities are by \cite{ku2}. This model represents
the best fit to the spectrum of MACHO Alert 95-30 obtained by \cite{sah}
and the available $V,R$ photometry. The model emergent spectrum in the optical
is shown in Figure 1a. At least seven TiO bands, CaH and CaI, can be identified
well on both the observed and synthesized spectra.
\begin{figure}
\plottwo{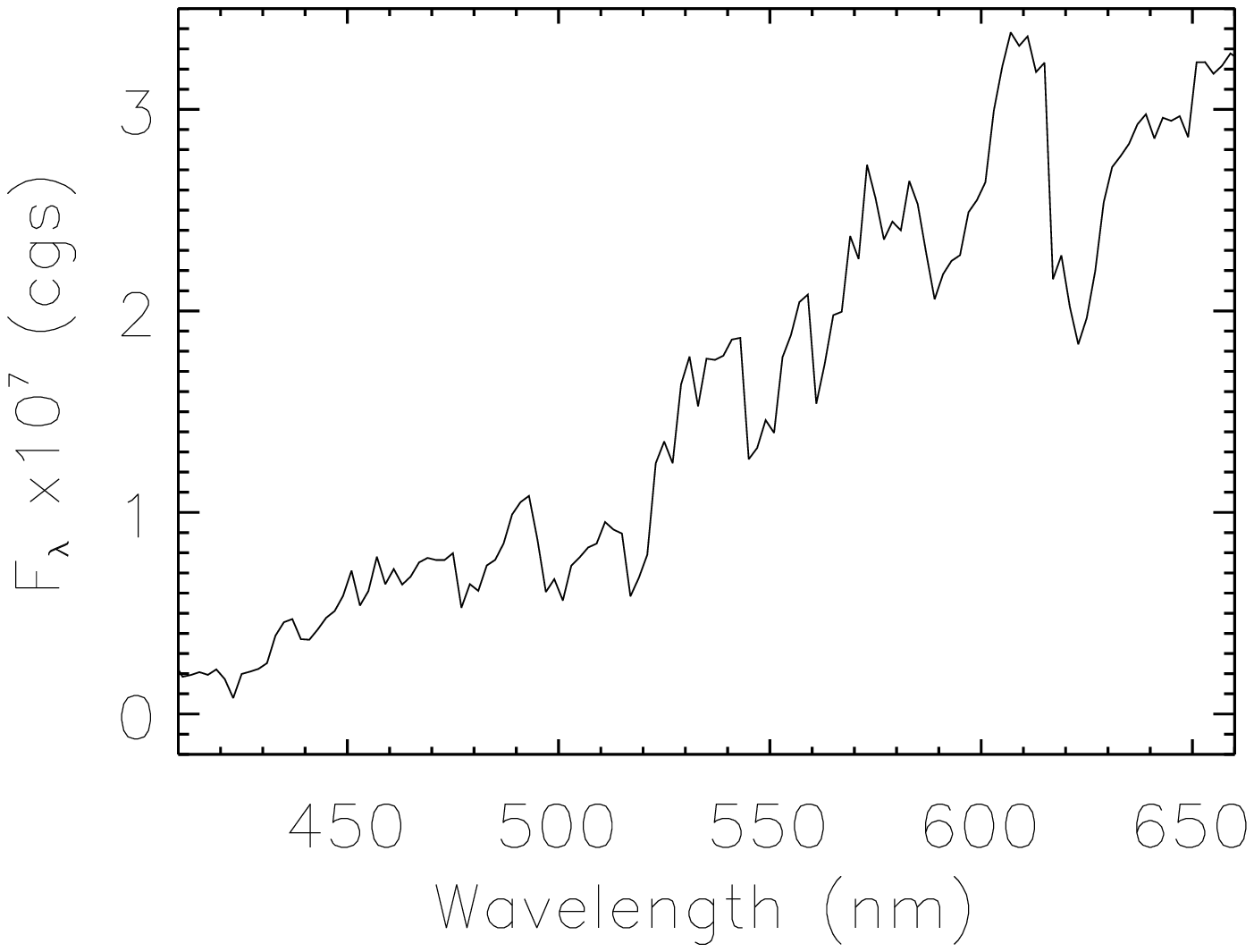}{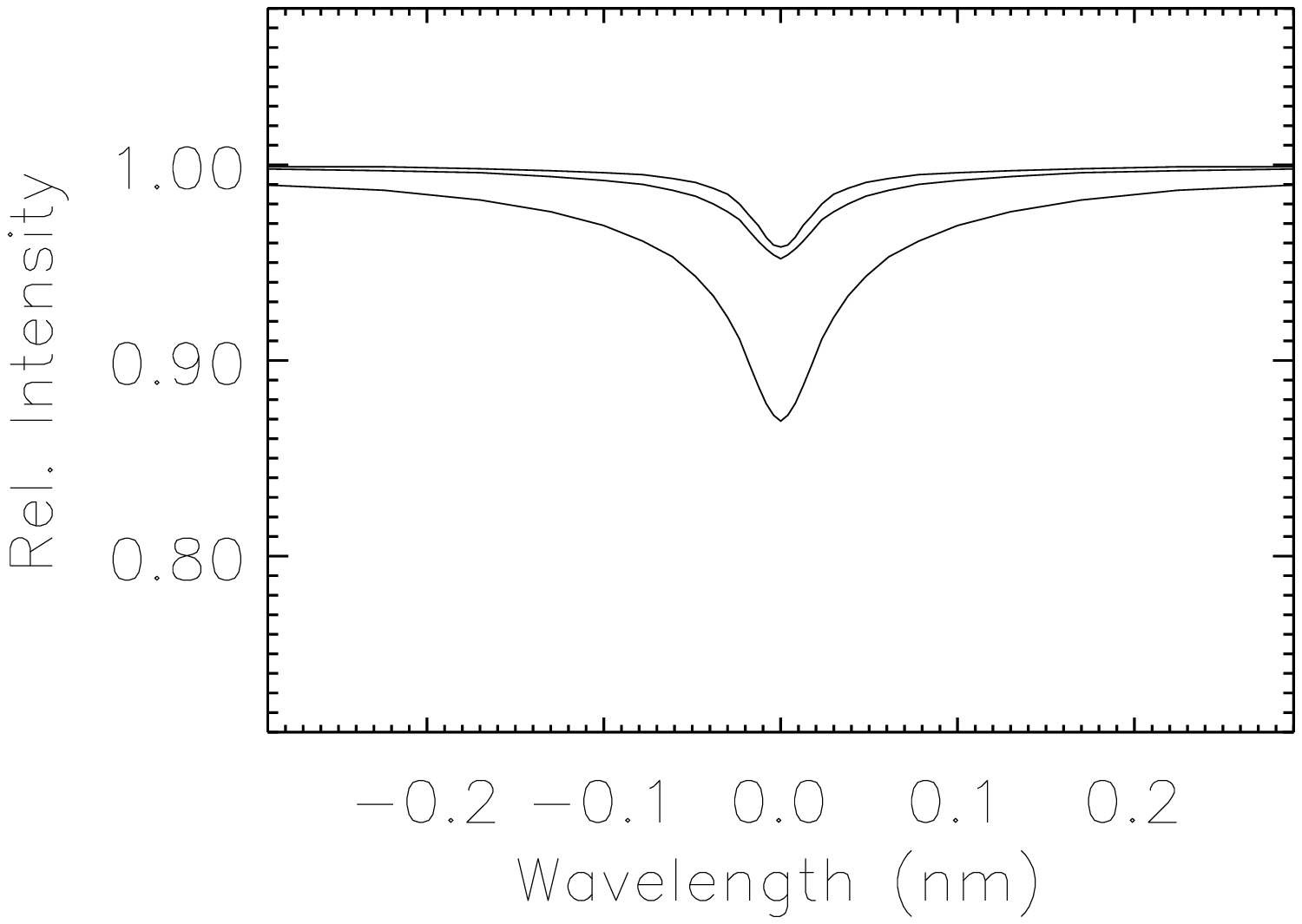}
\caption{The computed optical spectrum for the MACHO 95-30 source. a) The
absorption features at 476.2, 495.5, 516.8, 545, 588, \& 620 $nm$ are due to
TiO. 
b) Three profiles of the weak hydrogen H$\beta$ line from disk center at
$\mu$=0.98 (bottom), through $\mu$=0.5,
to very near the limb at $\mu$=0.02 (top).}
\end{figure}

\section{Stellar surfaces $-$ center-to-limb variations}
\subsection{Continuum}
Stellar disks are projected hemispheres, which implies an axisymmetric
variation of the intensity with position as it maps into a variation with
depth. In general, the emergent continuum radiation at the center of the disk
is formed deeper than the radiation we see near the limb. Over a very wide
spectral range for stars of different temperature, the center-to-limb variation
of the contunuum emission is manifested as limb darkening. The amount of limb
darkening is a function of wavelength. It has been included in calculations
of finite-size effects in microlensing \cite{wm,val6}.

An example of limb darkening in the $V$ and $R$ passbands and the
resulting microlensing light curves is given here in \cite{hey}.
The curves were calculated for the event MACHO Alert 95-30 with the stellar
model described above.

\subsection{Spectral lines}
Unlike continuum radiation, spectral lines provide a larger choice of
center-to-limb variations. In the majority of stars we deal with absorption
lines of different strength, arising from atomic or molecular bound-bound
transitions. Most moderately strong and weak lines diminish towards the limb
(Figure 1b). This variation could be observed in a microlensing event
by measuring the total equivalent widths of spectral lines on medium-resolution
spectra with high S/N. The magnitude of the effect is similar to that of
limb darkening, but involves a very narrow range of wavelengths and
hence a larger expense in the collection of photons.
\begin{figure}
\plottwo{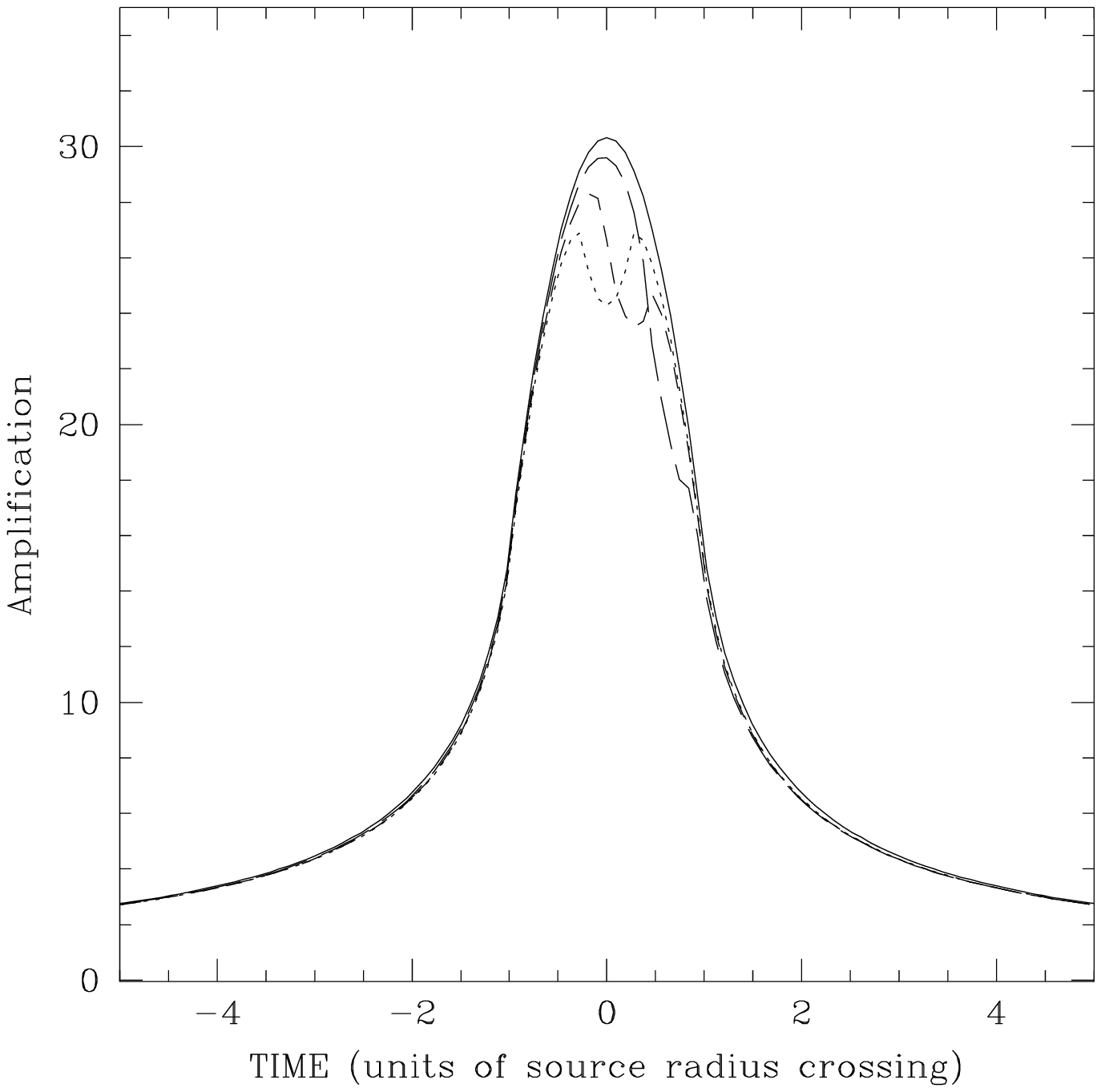}{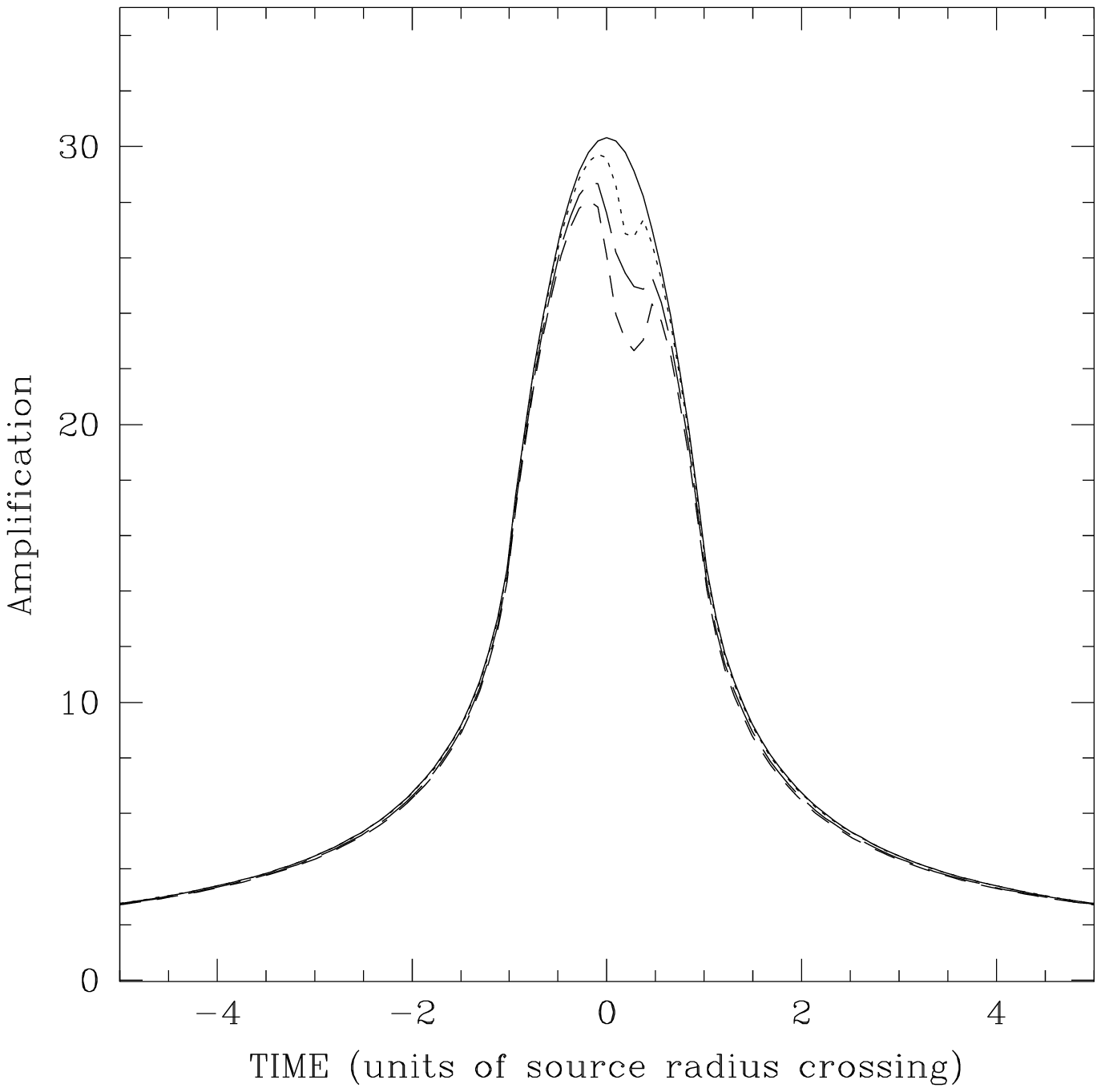}
\caption{Lightcurves for a red giant source with a dark spot. a) Three 
different positions of a spot with temperature difference (contrast)
$\Delta$T=500~K.
b) Different spot contrasts: $\Delta$T=250 \& 1000~K (dashed); and a twice
smaller spot (10\% of source radius) at $\Delta$T=500~K (dotted).}
\end{figure}

Therefore it is advantageous to look for spectral lines which have an
exceptionally strong center-to-limb variation. Obviously these will vary
by species and by type of star. A good choice would be lines that remain
optically thick at the limb and hence appear strongly in emission there.
Known as resonant line scattering, this effect is observed in the Sun
and should be common to cool stars in spectral lines like Ca~II~K. The
application of resonant line scattering in Ca~II to microlensing is
discussed in detail by \cite{ls}, who calculated light
curves and showed that it should be possible to observe the effect.
In stars of very low surface temperature, like the giant from
MACHO Alert 95-30, resonant line scattering may be observed in the
optical bands of TiO.

Resonant line scattering is greatly enchanced in the atmospheres of
stars which are extended (compared to our Sun) $-$ bright giants and
supergiants. So is the effect of limb polarization, discussed here by
\cite{col}. Polarimetric measurements, although more difficult,
would be a valuable probe of the stellar atmospheres of such stars.

\section{Stellar surfaces $-$ spots and other features}
Apart from having a venerable history (\cite{sch} and references
therein), the question of small-scale surface structure in normal stars
is very important for stellar modeling. Direct interferometric evidence
is scarce and inconclusive \cite{ben,hum}.
Indirect evidence, such as Doppler imaging and photometry, is
limited to specific types of stars. A very complete survey
of photometric evidence for stellar spots comes from the OGLE microlensing
survey of bulge giants \cite{ud5}.

In \cite{hey} we examined the effect of a stellar spot for a range of sizes and
contrasts ($\Delta$T) on the microlensing curve observed in an
optical passband for an event similar to MACHO Alert 95-30. Spots which 
are larger than 10\% of the radius have strong effect on the light
curve, and should be easy to detect (Figure 2). Such spots may be common
on red giants and could thus complicate the interpretation of light
curves which may be distorted due to a lens companion or planet.
To distinguish between spots on the source and companions of the lens,
it would be sufficient to have photometry in two or more
different spectral bands, as spot contrast is usually wavelength
dependent.

To maximize the search for spots and active regions on the surfaces of
microlensed stars one should again resort to the use of sensitive
spectral lines. Such techniques are widely known and used in the study
of the Sun $-$ one example is the bandhead of the CH radical at
430.5 $nm$, which provides very high contrast to surface structure
\cite{ber}. This will again require spectroscopy or very
narrow-band photometry of the microlensing event, but could be very
rewarding. The most practical approach is to microlens the entire
synthesized optical spectrum of the source in alerts similar to MACHO
95-30 and look for the most sensitive spectral features in advance of
the observations. The code described in \cite{hey} is
fast enough to allow this for the tens of thousands of frequencies
involved.

%
\end{document}